# Optimized Generation of Data-path from C Codes for FPGAs

Zhi Guo Betul Buyukkurt Walid Najjar
*University of California Riverside*
*{zguo, abuyukku, najjar}@cs.ucr.edu*

Kees Vissers
*Xilinx Corp.*
*kees.vissers@xilinx.com*

## Abstract

*FPGAs, as computing devices, offer significant speedup over microprocessors. Furthermore, their configurability offers an advantage over traditional ASICs. However, they do not yet enjoy high-level language programmability, as microprocessors do. This has become the main obstacle for their wider acceptance by application designers.*

*ROCCC is a compiler designed to generate circuits from C source code to execute on FPGAs, more specifically on CSoCs. It generates RTL level HDLs from frequently executing kernels in an application. In this paper, we describe ROCCC's system overview and focus on its data path generation. We compare the performance of ROCCC-generated VHDL code with that of Xilinx IPs. The synthesis result shows that ROCCC-generated circuit takes around 2x ~ 3x area and runs at comparable clock rate.*

## 1. Introduction

Continued increases in integrated circuit chip capacity have led to the recent introduction of Configurable System-on-a-Chip (CSoC), which has one or more microprocessors integrated with a field-programmable gate array (FPGA) as well as memory blocks on a single chip. In these platforms both the FPGA fabric, as well as the embedded microprocessors are essentially programmed using software. The earliest example is the Triscend E5 followed by the Triscend A7 [1], the Altera Excalibur [2], and Xilinx Virtex II Pro [3]. The capabilities of these platforms span a wide range with the Triscend A7 at the low end and the Xilinx Virtex II Pro 2VP125 at the high-end. These amazing computing devices have the flexibility of software and have been shown to achieve very large speedups, ranging from 10x to 100x, over microprocessors for a variety of applications including image and signal processing [4][5][6]. Such speedups come from large-scale parallelism made possible by high-capacity FPGAs, as well as from customized circuit design. The main problem standing in the way of wider acceptance of CSoC platforms is their programmability. Application developers must have an extensive hardware expertise, in addition to their application area expertise, to develop efficient designs. Presently, most CSoCs are programmed manually. The main drawback of this approach is that it is very labor intensive and requires large design times. Some commercial effort in programming FPGAs have been proposed by companies such as Synopsys [7] and Tensillica [8]. Their focus is on moving simple loops to hardware or on instruction-set extension.

Optimizing compilers for traditional processors have benefited from several decades of extensive research that has led to extremely powerful tools. Similarly, electronic design automation (EDA) tools have also benefited from several decades of research and development leading to powerful tools that can translate VHDL and Verilog code, and recently SystemC [9] code, into relatively efficient circuits. However, little work has been done to combine these two approaches. In other words, work is still needed to compile a high-level language program, based on C/C++/Java, with software level optimizations with the intent of generating a hardware circuit. Obviously, it is neither practical nor desirable to translate the whole program into hardware. It is therefore imperative to provide the programmer with tools that would help in identifying which code segments ought to be mapped to hardware as well as the cost and benefit tradeoffs implied.

Compiling to CSoCs and FPGAs in general is challenging. Traditional CPUs, including VLIW, have a fixed hardware platform. Their architectural features may or may not be exposed to the compiler. FPGAs, on the other hand, are completely amorphous. The task of an FPGA compiler is to generate both the hardware (data path) and the sequence of operations (control flow). This lack of architectural structure, however, presents a number of advantages. (1) The parallelism is very high and limited only by the size of the FPGA device or by the data memory bandwidth. (2) On-chip storage can be configured at will: registers are created by the compiler and distributed throughout the data path where needed, thereby increasing data reuse and reducing re-computations or accesses to memory. (3) Circuit customization: the data path and sequence controller are tailored to the specific computation being mapped to hardware. Examples include customized data bit-width and pipelining.

The objective of the ROCCC (Riverside Optimizing Configurable Computing Compiler) project is to design a high-level language compiler targeting CSoC. It takes high-level code, such as C or FORTRAN as input and generates RTL VHDL code for the FPGA and C code for the CPU. In this paper we describe the overall structure of the compiler and emphasize the data path generation component. We compare the clock speed and area of automatically generated circuits to a number of IP codes available on the Xilinx web site. The results show that the speed is within 10% while the area is larger by a factor of 2 to 3. The work in [25][26] has compared generated code with hand written VHDL. Both have shown a factor of 2 on the performance decrease of the generated code in area and clock rate. ROCCC is built upon the knowledge acquired from SA-C and Streams-C. We experimentally show that the resultant VHDL is much closer to the handwritten one.

The rest of this paper is organized as follows. The ROCCC compiler is introduced in section 2. Related work is discussed in section 3. Section 4 presents ROCCC compiler RTL code generation for the controller, the buffer and the data path. Experimental results are reported in section 5. Section 6 concludes the paper.

## 2. ROCCC System Overview

Figure 1 shows the overview of the ROCCC compiler. The profiling tool set has been described in a prior publication [10]. It identifies the frequently executing code kernels in a given application. ROCCC's objective is to compile these kernels to HDL code, which is synthesized using commercial tools.

The ROCCC system is built using SUIF [11] and Machine-SUIF [12] platforms. SUIF IRs (intermediate representations) provide abundant information about loop statements and array accesses. ROCCC performs loop level optimizations on SUIF IRs. Loop unrolling for FPGAs requires compile time area estimation. The work reported in [13] shows that in less than one millisecond and within 5% accuracy compile time area estimation can be achieved. Information to generate high-level units, such as controllers and buffers, is also extracted from SUIF IRs.

Machine-SUIF analysis and optimization passes, such as Control Flow Graph (CFG) library [14], Data Flow Analysis library [15] and Static Single Assignment library

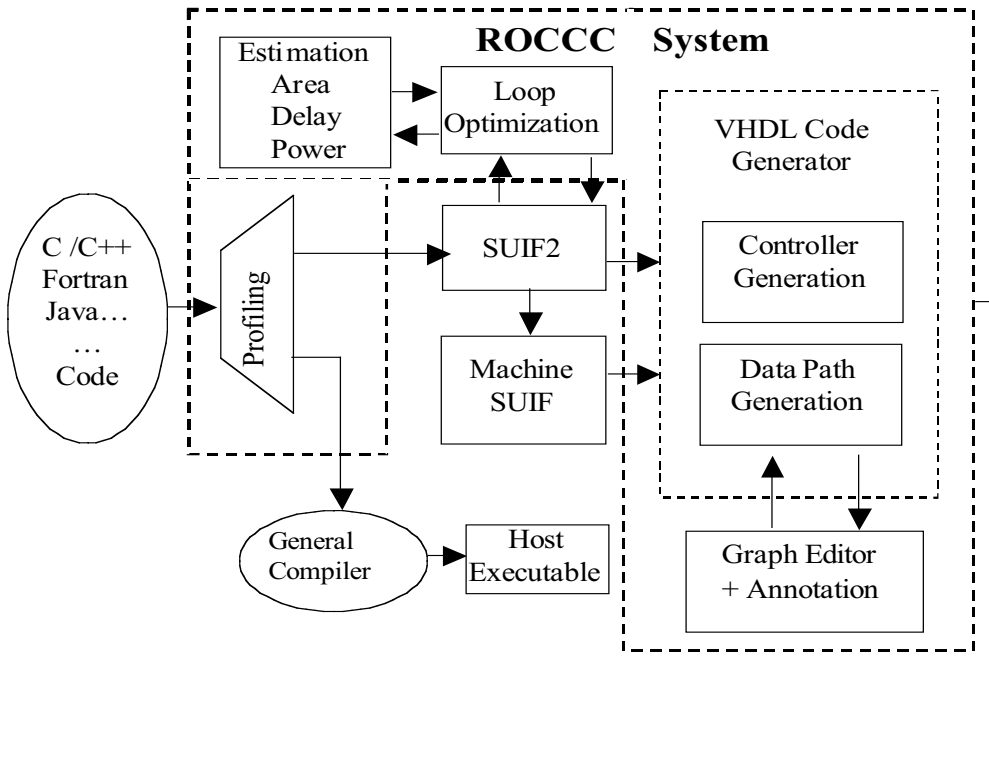

**Figure 1 - ROCCC System Overview**

[16], are used to generate the data path.

ROCCC's conventional optimizations include constant folding, loop unrolling, etc. Full loop unrolling converts a *for-loop* with constant bounds into a non-iterative block of code and therefore eliminates the loop controller. In addition to these conventional optimizations, at loop level ROCCC performs FPGA-specific optimizations, such as loop strip-mining, loop fusion, etc. At storage level and circuit level, ROCCC's optimizations are closely related with HDL code generation and are discussed in section 4.

The restrictions on the C code that can be accepted by the ROCCC compiler, for mapping on an FPGA fabric, include no recursion, no usage of pointers that cannot be statically unaliased. Function calls will either be inlined or whenever feasible made into a lookup table.

## 3. Related Works

Many projects, employing various approaches, have worked on translating high-level languages into hardware. SystemC [20] is designed to provide roughly the same expressive functionality of VHDL or Verilog and is suitable to designing software-hardware synchronized systems. Handle-C [21], as a low level hardware/software construction language with C syntax, supports behavioral descriptions and uses CSP-style (Communicating Sequential Processes) communication model.

SA-C [22] is a single-assignment high-level synthesizable language. Because of special constructs specific to SA-C (such as window constructs) and its functional nature, its compiler can easily exploit data reuse for window operations. SA-C uses pre-existing parameterized VHDL library routines to perform code generation in a way that requires a number of control signals between components, and thereby involves extra clock cycles and delay. Our compiler avoids spending clock cycles on handshaking by focusing more on the compile-time analysis. It takes a subset of C as input and does not involve any non-C syntax.

Streams-C [23] relies on the CSP model for communication between processes, both hardware and software. Streams-C can meet relatively high-density control requirements. However, it does not support accesses to two-dimension arrays and therefore image processing applications, including video processing, must be mapped manually. This makes it very awkward to efficiently support algorithms that rely on sliding windows. For one-dimension input data vector, such as a one-dimension FIR filter, Streams-C programmers need to manually write data reuse in the input C code in order to make sure that a data value is retrieved only once from external memory.

SPARK [24] is another C to VHDL compiler. Its transformations include loop unrolling, common sub-expression elimination, copy propagation, dead code elimination, loop-invariant code motion etc. SPARK does not support multi-dimension array accesses.

## 4. The ROCCC Compiler

ROCCC targets high computational density, low control density applications. Figure 2 shows the execution model. An engine moves the data from off-chip to a BRAM storage. The compiler-generated circuit accesses the arrays in BRAM and stores the output data into another BRAM, from which an engine retrieves data into the off-chip memory. Inside the compiler-generated circuit, the data path is fully pipelined. The controllers and buffers are in charge of feeding input data and retrieving output data to and from the data path.

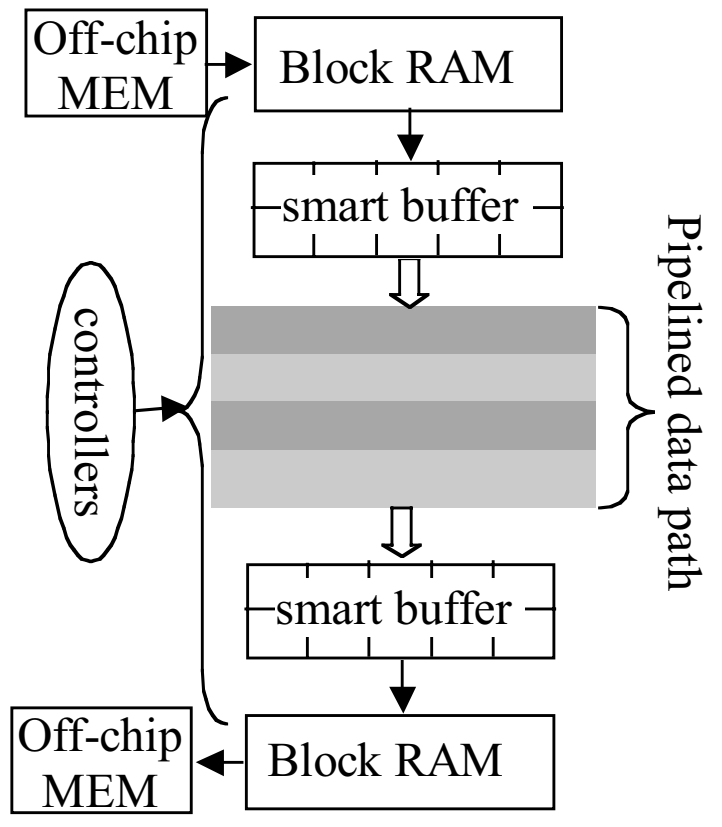

**Figure 2 - The Execution Model**

### 4.1 Controller and Buffers

ROCCC's scalar replacement transformation converts, for instance, the segment in Figure 3 (a) into the segment in Figure 3 (b). We can see that scalar replacement isolates memory access from calculation. The highlighted region of code is exported in the form of Figure 3 (c) and goes to the data path generator. At the same time, the loop statement and memory load/store code are used to generate the controllers and buffers. The controllers include address generators, which export a series of memory addresses according to the memory access pattern, and a higher-level controller, which controls the address generators. They are all implemented as pre-existing parameterized FSMs (finite state machine) in a VHDL library.

One of the major reasons that account for FPGA's speedup over general-purpose processor is that FPGA is capable of providing optimized I/O interface between data path and memory units [17]. For example, each iteration of the *for-loop* in Figure 3 (a) is essentially an operator on a window of five consecutive array elements. The window slides on the array. Two adjacent windows have four input data in common and only one new input data per window/iteration. ROCCC, as a high-level synthesis compiler, uses the knowledge of memory access pattern from the input code, such as the code shown in Figure 3 (b), to automatically generates an intelligent buffer, called *smart buffer*, based on the bus size, window size, data size and sliding-window stride. This buffer unit is able to reuse live input data, clean unused data and export the present valid input data set (the 5-data window in Figure 3 (b)) to the data path [18].

```
for (i=0; i<N; i=i+1) {
  C[i] = 3*A[i] + 5*A[i+1] + 7*A[i+2] + 9*A[i+3] – A[i+4]; }
```
(a)

```
for (i=0; i<17; i=i+1) {
  A0 = A[i];      A1 = A[i+1];   A2 = A[i+2];
  A3 = A[i+3];   A4 = A[i+4];
  Tmp0 = 3*A0 + 5*A1 + 7*A2 + 9*A3 - A4;
  C[i] = Tmp0;  }
```
(b)

```
void main_df(int A0,int A1,int A2,int A3,int A4,int* Tmp0)
{
  *Tmp0 = 3*A0 + 5*A1 + 7*A2 + 9*A3 - A4;
  return;  }
```
(c)

(a) – A 5-tap FIR in original C code
(b) – The FIR after scalar replacement
(c) – The FIR C code fed into the data path generator

**Figure 3 - A 5-tap FIR in C**

### 4.2 Data Path Generation

Before building the data path, a few preparation passes are done both at the front-end and back-end. Then, ROCCC's back-end passes perform the analysis, optimization and data path generation.

#### 4.2.1 Preparation Passes

ROCCC uses Machine-SUIF virtual machine (SUIFvm) [19] intermediate representation as the back-end IR. The original SUIFvm assembly-like instructions, by themselves, cannot completely cover HDLs' hardware description functionality. On the other side, the front-end analysis may assist and simplify the data path generation at back-end. Besides back-end data flow analysis, ROCCC performs high-level data flow analysis at front-end and the analysis information is transferred through pre-defined macros to assists back-end hardware generation.

Figure 4 (b) shows an accumulator after applying scalar replacement in C. The variable *sum* is detected as a feedback signal. Figure 4 (c) shows the resultant segment in C, in which macro ROCCC_load_prev() and macro ROCCC_store2next() annotate the signal feedback.

After applying scalar replacement and front-end dataflow analysis, the function that describes the scalar computing, like the codes shown in Figure 3 (c) or Figure 4 (c), is fed into Machine-SUIF. ROCCC performs circuit level optimizations and eventually generates data path on a modified version of the Machine-SUIF virtual machine (SUIFvm) [19] intermediate representation.

Before fed to ROCCC's passes, the virtual machine IR first undergoes Machine-SUIF Static Single Assignment and Control Flow Graph transformations. At this point, control flow graph information is visible and every virtual register is assigned only once.

The preserved macros are converted into ROCCC-specific opcodes. For example, ROCCC_load_prev() and ROCCC_store2next() in Figure 4 (c) are converted into instructions with opcode *LPR* (load previous) and *SNX* (store next), respectively. We are working on supporting bit

```
int sum = 0;
for ( i = 0; i < 32; i++) {
sum = sum + A[i];
}
```
(a)

```
int sum = 0;
for ( i = 0; i < 32; i++) {
main_Tmp0 = A[i];
sum = sum + main_Tmp0;
}
```
(b)

```
int sum = 0;
void main_dp(int main_Tmp0, int* main_Tmp1) {
int main_dp_Tmp2;
main_dp_Tmp2 = ROCCC_load_prev(sum) + main_Tmp0;
ROCCC_store2next(sum, main_dp_Tmp2);
*main_Tmp1 = sum;
}
```
(c)

(a) – An accumulator in original C code
(b) – The accumulator after scalar replacement
(c) – The C code fed into data path generator after detecting feedback variable and adding preserved macros

**Figure 4 - An Accumulator in C**

manipulation macros, which are the lack of high-level languages.

### 4.2.2 Data Path Building

Each instruction that goes to hardware is assigned a location in the data path. We add new fields into Machine-SUIF IR to record the location of each arithmetic, logic or register copying instruction's location. For example, Figure 6 shows the data path for the C code list in Figure 5. We maximize instruction level parallelism. All the input and output operands are copied to the entry or exit of the data flow, respectively. A virtual register's definition and reference should be adjoining in the data flow. If not, extra register copying instructions are added to satisfy so.

The compiler first builds data path for each non-null node in the CFG, as node 1 through node 4 shown in Figure 6. To parallelize alternative branches, the compiler adds a new *mux* node between alternative branch nodes and their common successor node, for instance, node 7 in Figure 6. A new *pipe* node (node 6 in Figure 6, for instance) is added

```
void if_else(int x1, int x2, int* x3, int* x4) {
   int a,c;
   c = x1 - x2; /*node 1*/
   if(c < x2)
      a = x1*x1 ; /*node 2*/
   else
      a = x1 * x2 + 3; /*node 3*/
   c = c - a;  }
   *x3 = c;    } node 4
   *x4 = a;    }
   return;}    }
```

**Figure 5 - An Alternative Branch in C**

* The pointers are only used to indicate multiple return values. ROCCC does not support pointers

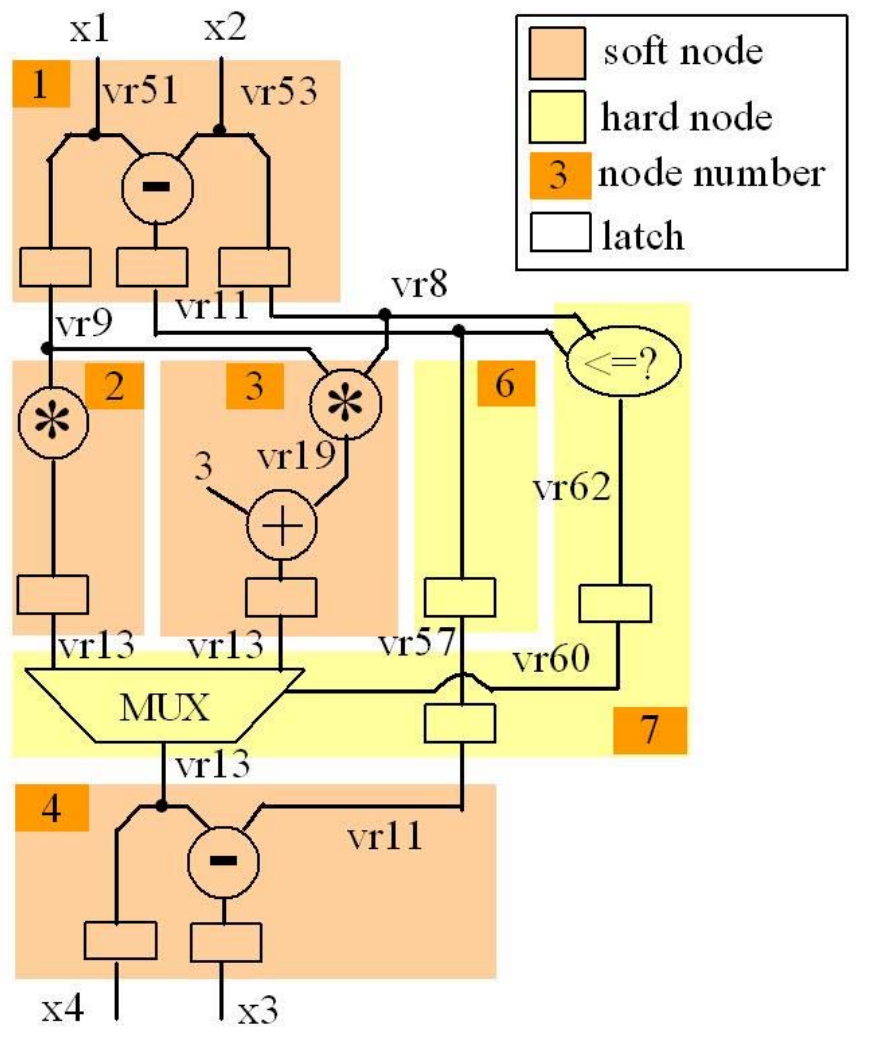

**Figure 6 - The Alternative Branch Data Path**

to copy live variables from alternative branches' parent node to their common successor node.

In Figure 6, node 6 and 7 are called hard nodes since they only appear in hardware and have no equivalence in software. Nodes 1 through 4 are thereby called soft nodes. Notice that if we only consider soft node, vr11 in node 4 is vr11 in node 1, the same case as of vr13. Therefore, the soft nodes, by themselves, will have the same behavior on a CPU compared with the whole data path on a FPGA.

### 4.2.3 Data Path Pipeline

ROCCC automatically places latches in a data path to pipeline it. The latch location in a node is decided based on the delay estimation of instructions, which is beyond this paper's scope.

The latch location also satisfies special opcodes' requirements. For example, *SNX* instruction must have a latch to store the feedback signal to the corresponding *LPR* instruction. Figure 7 shows the data path of Figure 4 (c).

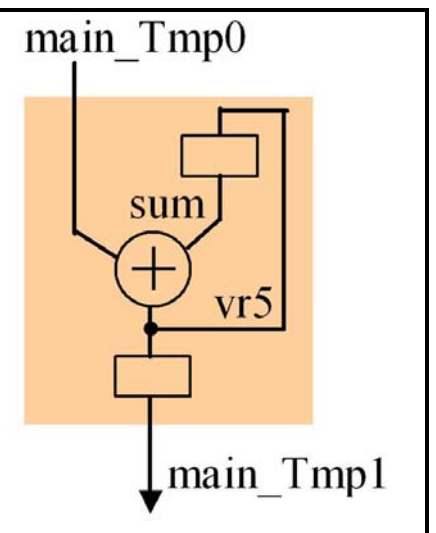

**Figure 7 - The Accumulator Data Path**

After data path pipelining, each pipeline stage is an instance of single iteration in the *for-loop* body.

### 4.2.4 VHDL Code Generation

ROCCC generates one VHDL component for each CFG node that goes to hardware. In a node, every virtual register is single assigned and is converted into wires in hardware. All arithmetic opcodes in SUIFvm have corresponding functionality in IEEE 1076.3 VHDL with the exception of division. Arithmetic, logic and copying instructions become combinational or sequential VHDL statement according to whether the instruction needs latched or not. A *LUT* instruction invokes an instantiation of a lookup table component. If the lookup table is a pre-existing one, such as a *cos* lookup table, the compiler automatically includes the existing component. Otherwise, for example, if a user wants to have a probability distribution function lookup table, the compiler instantiates the lookup table as a regular *ROM* IP core unit in the VHDL code. The only thing the user needs to do is to edit a pure text initialization file, which defines the lookup table's content.

**Table 1:** A comparison of hardware performance from Xilinx IPs and ROCCC-generated VHDL code.
(*The wavelet engine is not from the Xilinx IP, it is written in VHDL)

| | Xilinx IP | | ROCCC | | | |
|---|---|---|---|---|---|---|
| Example | Clock (MHz) | Area (slice) | Clock (MHz) | Area (slice) | %Clock | %Area |
| bit_correlator | 212 | 9 | 144 | 19 | 0.679 | 2.11 |
| mul_acc | 238 | 18 | 238 | 59 | 1.00 | 3.28 |
| udiv | 216 | 144 | 272 | 495 | 1.26 | 3.44 |
| square root | 167 | 585 | 220 | 1199 | 1.32 | 2.05 |
| cos | 170 | 150 | 170 | 150 | 1.00 | 1.00 |
| Arbitrary LUT | 170 | 549 | 170 | 549 | 1.00 | 1.00 |
| FIR | 185 | 270 | 194 | 293 | 1.05 | 1.09 |
| DCT | 181 | 412 | 133 | 724 | 0.735 | 1.76 |
| Wavelet* | 104 | 1464 | 101 | 2415 | 0.971 | 1.65 |

By adding more data type in Machine-SUIF, ROCCC supports any signed and unsigned integer type up to 32 bit. The compiler infers the inner signals' bit size automatically.

## 5. Experimental Results

We compare the hardware performance generated from Xilinx IP cores and ROCCC-generated VHDL code. We use Xilinx ISE 5.1i and IP core 5.1i. All the Xilinx IP cores and ROCCC-generated VHDL code are synthesized targeting a Xilinx Virtex-II xc2v2000-5 FPGA.

All the benchmarks in Table 1 are picked from Xilinx IP core except the wavelet engine. The input and output variables of ROCCC equivalents have the same bit sizes as that of the IP cores.

*Bit_correlator* counts the number of bits of an 8-bit input data that are the same as of a constant mask. *Mul_acc* is a multiplier-accumulator, whose input variables are a pair of 12-bit data. *Udiv* is an 8-bit unsigned divider. *Square_root* calculates a 24-bit data's square root. *Cos*'s input is 10-bit, output, 16-bit. The arbitrary LUT has the same port size as that of *cos*. *FIR* is two 5-tap 8-bit constant coefficient finite impulse response filters, whose bus sizes are 16-bit. *DCT* is a one-dimension 8-data discrete cosine transform. The input data size and output data size are 8-bit and 19-bit, respectively. For Xilinx IP *FIR* and *DCT*, the multiplications with constants are implemented using distributed arithmetic technique, which performs multiplication with lookup-table based schemes. Therefore, we set the synthesis option *'multiplier style'* as *'LUT'* for the ROCCC-generated *DCT* and *FIR*.

The second and the third column of Table 1 show Xilinx IP cores' clock rate and device utilization and the forth and the fifth column show ROCCC's corresponding performance. %Clock is the percentage difference in clock rate of ROCCC-generated VHDL compared to Xilinx IP. %Area is the percentage difference in area of ROCCC-generated VHDL compared to Xilinx IP. *Bit-correlator, udiv* and *square root* consist of a number of bit manipulations. The C input, as a high-level code, is not good at describing bit operations and therefore is one of the major causes of the performance difference. Xilinx *mul_acc* IP has a control input signal *nd (new data)* whose Boolean value *true* indicates the present data is valid. In C code, we describe the equivalent behavior using *if-else* statement whose condition evaluates Boolean input *nd.* Thus, extra nodes and latches are added to support the alternative branch and take extra area. We used to convert this C code by multiplying *nd* with the new input data instead of using *if-else* statement. Though one more multiplier was used, the overall area and clock rate performance was better than the one listed in Table 1. Obviously, this is not compile level optimization. But at the same time, it shows one of high-level synthesis's advantages: ease to do algorithm level optimizations. In terms of lookup tables, ROCCC-generated VHDL code instantiates Xilinx IP cores. Therefore, they have exactly the same performance. In Xilinx Virtex-II, 10-bit-input-16-bit-output *cos/sin* lookup table stores only half wave, which is one of the reasons that this *cos/sin* lookup table utilizes less area compared with the arbitrary ROM lookup table with the same port size. *Fir* operates on an array. Basically, a 5-data window slides on the one-dimension array. ROCCC generates *smart buffer* to reuse the previous input data. The *FIR*'s data path consists of multipliers, adders/subtracters and no branch. ROCCC fits this type of algorithms and gets comparable performance with IP cores. Like *FIR*, *DCT* has high computational density and no branch. The throughput of Xilinx DCT IP is one output data per clock cycle, while ROCCC's throughput is eight output data per clock cycle. Therefore, though ROCCC-generated DCT runs at a lower speed (73.5%), the overall throughput of ROCCC-generated circuit is higher. Both ROCCC DCT and Xilinx IP DCT explore the symmetry within the cosine coefficients. The last row in Table 1 shows an implementation of a two-dimension (5, 3) wavelet transform engine, which is the standard lossless JPEG2000 compression transform. This wavelet transform engine includes the *address generator*, *smart buffer* and data path. The ROCCC-generated circuit is compared with a handwritten one.

We derive bit width only based on port size and opcodes. More aggressive bit narrowing, performed by users or/and the compiler, may reduce device utilization.

## 6. Conclusion

The reconfigurable computing paradigm is a powerful computing model that has a lot of potential for long-running or streaming applications that are somehow regular in nature. The main obstacle to its use is its programmability. Handwritten HDL code for large scale applications is not the most desirable approach. Automatic compiler generation of HDL code from high-level languages is very challenging.

The ROCCC compiler generates VHDL for reconfigurable computing from high-level languages, such as C or Fortran. ROCCC performs loop level, storage level and circuit level optimizations. In this paper we have mainly presented its data path generation. At front-end, the compiler performs high-level data flow analysis and transfers the analysis information through preserved macros. At back-end, the compiler explores low-level parallelism, pipelines data path and narrows inner signals' bit sizes. ROCCC supports lookup tables through automatically instantiating pre-existing lookup table IPs or ROM IPs.

We compared the performance of ROCCC-generated VHDL code with that of Xilinx IPs. The synthesis result shows that ROCCC-generated circuit takes around 2x ~ 3x area and runs at comparable clock rate. ROCCC performs better on high computational density examples than on high control density ones.